%% file: JSC_LES_ICOSAHOM04_Online_Version.tex
\documentclass[]{elsart}
\usepackage{longtable}
\usepackage{epsfig}
\usepackage{amssymb}
\usepackage{amsmath}
\usepackage{bm}
\usepackage{graphicx}
\usepackage{graphics}
\usepackage{color}
\usepackage{pstricks}
\usepackage{cite}
\def\bdm{\begin{displaymath}}
\def\enddm{\end{displaymath}}
\def\beq{\begin{equation}}
\def\endeq{\end{equation}}
\def\beqn{\begin{eqnarray}}
\def\endeqn{\end{eqnarray}}

\date{}
\newcommand{\dd}{\text{d}}

\newcommand{\Lmat}{\left(\begin{array}}
\newcommand{\Rmat}{\end{array}\right)}

\begin{document}

\begin{frontmatter}
\journal{\it J. Sci. Comput.}
\volume{27}
\issue{1}
\pubyear{2006}
\firstpage{151}
\lastpage{162}

\title{ \bf Large-eddy simulation of the lid-driven cubic cavity flow by the spectral element method} 
\author[EPFL]{Roland Bouffanais\corauthref{cor}},
\corauth[cor]{Corresponding author.}
\ead{roland.bouffanais@epfl.ch}
\author[EPFL]{Michel O. Deville}
\ead{michel.deville@epfl.ch}
\author[MCS]{Paul F. Fischer},
\ead{fischer@mcs.anl.gov}
\author[EPFL]{Emmanuel Leriche},
\ead{emmanuel.leriche@epfl.ch}
\author[EPFL]{Daniel Weill}
\address[EPFL]{Laboratory of Computational Engineering, \'Ecole Polytechnique F\'ed\'erale de Lausanne, STI--ISE--LIN Station 9, CH-1015 Lausanne, Switzerland}
\address[MCS]{MCS, Argonne National Laboratory, 9700 S. Cass Avenue, Argonne, IL 60439, USA}
\begin{abstract}
This paper presents the large-eddy simulation of the lid-driven cubic cavity flow by the spectral element method (SEM) using the dynamic model. Two spectral filtering techniques suitable for these simulations have been implemented. Numerical results for Reynolds number $\text{Re}=12'000$ are showing very good agreement with other experimental and DNS results found in the literature.

\end{abstract}
\end{frontmatter}

\section{Introduction}
Spectral element methods have been mainly applied to the direct numerical simulation (DNS) of fluid flow problems at low and moderate Reynolds (Re) numbers. With the advent of more powerful computers, especially through cluster technology, higher Re values seem to fall within the realm of feasibility. However, despite their high accuracy, spectral element methods are still far from reaching industrial applications that involve developed turbulence at Re values of the order of $10^6-10^7$. The reason for that dismal performance is that a resolved DNS including all scales from the gross structures to Kolmogorov scales, needs a number of degrees of freedom (dof) that grows like $\text{Re}^{9/4}$. Therefore with increasing Re, we have to increase the number of elements, $E$, and the degree, $N$, of the polynomial spaces. This places the computational load far out of the reach of present day computers. 

Large-eddy simulation (LES) represents an alternative to DNS insofar that it involves less dofs because the behaviour of the small scales are modeled. The LES methodology deals with high Re number unsteady flows by using coarse meshes in which the contributions of small sub-grid scale (SGS) structures are modeled while the large-scale structures are obtained by the computed flow dynamics. The reader is referred to the monograph by Sagaut \cite{sagaut03} for details. In this paper we will focus our attention on the dynamic model \cite{germ91,sagaut03,lesieur96} which will be evaluated by comparing LES numerical results and DNS on the three-dimensional cubic cavity flow.
\section{Mathematical modelling}
\subsection{The governing equations}
Large-scale quantities, designated by an ``overbar'', are obtained by a filtering process on the domain $\Omega$. Assuming the filter commutes with differentiation and applying the filter to the Navier--Stokes equations in divergence form for the nonlinear term, one obtains the following relations:
\beqn  
\frac {\partial {\bf \overline  v}}  {\partial t} + {\bm \nabla} \cdot {\bf \overline {v} \, \overline {v}} \,&=&\, -{\bm \nabla} {\overline p} +\nu {\bm \Delta} { \bf \overline v}- {\bm \nabla} \cdot \mbox{\boldmath $ \tau$}, \label{eq:filterNS}\\
{\bm \nabla} \cdot\, {\bf \overline v} & = & 0.\label{eq:LES} 
\endeqn 
Here, ${\bf \overline v}$ is the filtered velocity, $t$ denotes the time, $\overline p$ is the filtered pressure divided by the constant density, $\nu$ the kinematic viscosity. The symbols $\bm \nabla$ and $\bm \Delta$ represent the nabla and Laplacian operators, respectively. The SGS stress tensor $\mbox{\boldmath $\tau$}$ takes the small-scale effects into account and is given by
\beq
\mbox{\boldmath$\tau$}={\bf \overline {vv}}-{\bf \overline v} \, {\bf \overline v}. 
\label{eq:SGS}
\endeq 
\subsection{Smagorinksy and dynamic models}
The SGS Smagorinsky model (SM) \cite{smag63} uses the concept of turbulent viscosity and  assumes that the small scales are in equilibrium, balancing energy production and dissipation. This yields the following expression for the eddy-viscosity
\beq 
\nu_{T} = (C_{\text{S}}\overline \Delta)^2  |\overline S|, \label{eq:lesconst}
\endeq 
where $|{\bf \overline S}|=(2\overline S_{ij} \overline S_{ij})^{1/2}$ is the magnitude filtered strain-rate-tensor, $C_{\text{S}}$ is the Smagorinsky constant and $\overline \Delta$ the  filter width. The SM has several drawbacks. The most severe one is the constant value of $C_{\text{S}}$ during the computation which produces too much dissipation. Furthermore the SM does not provide the modeller with backscattering where kinetic energy is transferred from small scales to larger scales. 

The dynamic model (DM) proposed by Germano et al. \cite{germ91} overcomes the difficulty of constant $C_{\text{S}}$, by allowing it to become dependent of space and time. Now we have a dynamic constant $C_{\text{d}}=C_{\text{d}}({\bf x}, t)$. Let us introduce a test-filter length scale $\hat \Delta$ that is larger than the grid length scale $\overline \Delta$ (e.g. $\hat \Delta=2\overline \Delta$). Using the information provided by those two filters and assuming that in the inertial range of the turbulence energy spectrum, the statistical self-similarity applies, we can better determine the features of the SGS stress. With the test filter, the former LES equations (\ref{eq:filterNS}) yield a relation involving the sub-test-scale stress 
\beq
{\bf T}={\bf \widehat {\overline {vv}}}-{\bf \hat {\overline v}} \, {\bf \hat {\overline v}}. 
\label{eq:STS}
\endeq 
 We introduce the Germano identity to obtain the relation between ${\bf T}$ and the filtered $\mbox{\boldmath$\hat \tau$}$ such that
\beq
{\bf L}={\bf T}-\mbox{\boldmath$\hat \tau$}={\bf \widehat {\overline v \,  \overline v}}-{\bf \hat {\overline v} \, \hat {\overline v}}.
\label{eq:Ltensor}
\endeq
 We apply the eddy viscosity model to {\boldmath $\tau$} and ${\bf T}$ and we obtain using the self-similarity hypothesis for the constant $C_{\text{d}}$
\beq 
\text{\boldmath $\tau$} -\frac{1}{3} \text{tr}(\text{\boldmath $\tau$})\, {\bf I}=-2 C_{\text{d}} {\bf \overline {\Delta}}^2 |{\bf \overline S}| {\bf \overline S}= C_{\text{d}} \text{\boldmath $\beta$}, \label{eq:taueq}
\endeq 
and
\beq 
{\bf T} -\frac{1}{3} \text{tr}({\bf T})\, {\bf I}=-2C_{\text{d}}
 {\bf \hat {\overline \Delta}}^2|{\bf \hat {\overline S}}| {\bf \hat { \overline S}}= C_{\text{d}} \text{\boldmath $\alpha$}
, \label{eq:Teq}
\endeq 
where the symbol tr denotes the trace of the tensor. Inserting (\ref{eq:taueq}) and (\ref{eq:Teq}) in the deviatoric part ${\bf L}^\text{d}$ of $\bf L$ produces 
\beq
{\bf L}-\frac{1}{3}\text{tr}({\bf L}){\bf I}\equiv {\bf L}^\text{d}=C_\text{d} \text{\boldmath $\alpha$}- \widehat{C_{\text{d}}\text{\boldmath $\beta$}}.\label{eq:devG}
\endeq
Assuming that $C_{\text{d}}$ does not vary too much in space, one sets ${\widehat C_{\text{d}}} \approx C_{\text{d}}$ and one can deduce from a least square  minimization of the error related to (\ref{eq:devG}) (see \cite{lilly92,sagaut03}) that 
\beq
C_{\text{d}}=\frac{(\text{\boldmath $\alpha$}-\widehat{\text{\boldmath $\beta$}}):{\bf L}^\text{d}}{(\text{\boldmath $\alpha$}-\widehat{\text{\boldmath $\beta$}}):(\text{\boldmath $\alpha$}-\widehat{\text{\boldmath $\beta$}})}, \label{eq:Cdyn}
\endeq
where the notation $:$ is used for inner tensor product (double contraction).
\section{Numerical approximation}
\subsection{Space discretization}
The numerical approximation is obtained through a weak formulation of the Eqs. (\ref{eq:filterNS})--(\ref{eq:LES}) discretized using the Lagrange--Legendre approximation. The reader is referred to the monograph by Deville, Fischer and Mund \cite{dfm02} for details. The velocity and pressure are expressed in the ${\mathbb P}_N - {\mathbb P}_{N-2}$ spaces where ${\mathbb P}_N$ is the set of polynomials of degree $\leq N$ in each space direction. This approximation avoids the presence of spurious pressure modes as it was proved by Maday and Patera \cite{maday89}. The quadrature rules involved in the weak formulation define a Gauss--Lobatto--Legendre (GLL) grid for the velocity nodes and a Gauss--Legendre grid (GL) for the nodal pressures. 

\hspace{-0.15ex}Borrowing the notation from \cite{dfm02}, the semi-discrete filtered Navier--Stokes equations resulting from the spectral element discretization are
\beqn
{\bf M} \frac {\dd \underline {\bf \overline v}}{\dd t}+ {\bf C \underline {\bf \overline v}} +\nu \, {\bf K}{\underline {\bf \overline v}}-{\bf D}^T {\underline{\overline p}}+{\bf D}{\underline {\mbox{\boldmath $\tau$}}}&=&0,\label{eq:odes}\\
-{\bf D}{\underline {\bf \overline v}}&=&0\label{eq:constr}.
\endeqn
The diagonal mass matrix ${\bf M}$ is composed of $d$ blocks, the mass matrices $M$, with $d=3$ for the three-dimensional cavity problem. The supervector ${\underline {\bf \overline v}}$ contains all the nodal velocity components while ${\underline {\overline p}}$ is made of all nodal pressures. The matrices ${\bf K}, {\bf D}^T, {\bf D}$ are the discrete Laplacian, gradient and divergence operators, respectively. The matrix operator ${\bf C}$ represents the action of the non linear advection term written in convective form ${\bf \overline v}\cdot {\bm \nabla}$, on the velocity field and depends on $\underline {\bf \overline v}$ itself. The spatial discretization leads to a set of non linear ordinary differential equations  (\ref{eq:odes}) subject to the incompressibility condition (\ref{eq:constr}).
\subsection{Time integration}
As the LES viscosity is not invariant, we modify the standard time integration scheme in such a way that this space varying viscosity be handled explicitly as this was done e.g. in \cite{karam99, black03}. Let us define the effective viscosity as
\beq
\nu_{\text{\scriptsize eff}}=\nu +\nu_T=\nu_{\text{\scriptsize cst}}+(\nu_{\text{\scriptsize eff}}-\nu_{\text{\scriptsize cst}})
\endeq
where $\nu_{\text{\scriptsize cst}}$ is the sum of $\nu$ and the spatial average of $\nu_T$ over the computational domain, being by construction constant in space but not in time.
The filtered Navier--Stokes equations become
\beqn
{\bf M} \frac {\dd \underline {\bf \overline v}}{\dd t} +\nu_{\text{\scriptsize cst}} \, {\bf K}{\underline {\bf \overline v}} -{\bf D}^T {\underline{\overline p}}&=&-{\bf C \underline {\bf \overline v}}+2{\bf D}(\nu_{\text{\scriptsize eff}}-\nu_{\text{\scriptsize cst}}){\overline {\bf \underline S}}\label{eq:odefin}\\
-{\bf D}{\underline {\bf \overline v}}&=&0.
\endeqn
The viscous linear term and the pressure are implicitly integrated by a backward differentiation formula of order $2$ (BDF2) while the terms in the right-hand side of Eq. (\ref{eq:odefin}) are computed by
a second order extrapolation method (EX2). The explicit viscous part $2{\bf D}(\nu_{\text{\scriptsize eff}}-\nu_{\text{\scriptsize cst}}){\overline {\bf \underline S}}$ leads to a stability condition such that $(\nu_{\text{\scriptsize eff}}-\nu_{\text{\scriptsize cst}})\Delta t \leq C/N^4$ while the CFL condition restricts the time step such that $\overline u_{\text{max}} \Delta t \leq C/N^2$. It would seem that the viscous restriction is more severe than the convective one. However the magnitude of the term $2{\bf D}(\nu_{\text{\scriptsize eff}}-\nu_{\text{\scriptsize cst}}){\overline {\bf \underline S}}$ is far smaller than the one of the convective term. Therefore the stability is indeed enforced by the CFL limit.

The implicit part is solved by a generalized block LU decomposition, using a standard fractional-step method with pressure correction which may be preconditioned by various algorithms.
\section{Filtering}
As spectral elements offer high accuracy for the flow at hand, we construct the filters using two spectral techniques. The first one is a nodal filter acting in physical space on the nodal velocity components (and pressure) to stabilize the computations. The second method is designed as a modal filter and is carried out element-wise in spectral space and corresponds to the convolution kernel of the LES filtering. 
\subsection{Nodal filter}
The nodal filter is due to Mullen and Fischer \cite{mul99} and is adequately suited to parallel spectral element computation. Introducing $h_{N,j}, j=0,\ldots,N$ the set of Lagrange-Legendre interpolant polynomials of degree $N$ on the GLL grid nodes $\xi_{N,k}, k=0,\ldots,N$, the rectangular matrix operator $I^M_N$ of size $(M+1)\times(N+1)$ is such that
\beq
(I^M_N)_{ij}=h_{N,j}(\xi_{M,i}).
\endeq
Therefore, the  matrix operator of order $N-1$
\beq
\Pi_{N-1}=I^N_{N-1}\,I^{N-1}_N
\endeq
interpolates on the GLL grid of degree $N-1$ a function defined on the GLL grid of degree $N$ and transfers it back to the original grid. By this process, one can show that one gets rid of the highest modes of the polynomial representation. The $1D$ filter is given by the relation
\beq
{\overline v}=[\alpha \Pi_{N-1}+(1-\alpha)I^N_N]v.
\endeq
The LES version of the filter sets $\alpha=1$ and is given by
\beq
{\overline v}=I^N_M\,I^M_Nv,
\endeq
where $M$ is equal to $N-2$ or $N-3$. The extension to three-dimensional problems results easily from the matrix tensor product properties of the filter.
\subsection{Modal filter}
The variable $v$ aside its Lagrange-Legendre representation may also be approximated by a modal basis that was first proposed in the $p$ version of the finite elements. That basis was used by Boyd \cite{boyd98} as a filter technique and is built up on the reference parent element as
\beq
\phi_0=\frac{1-\xi}{2},\, \phi_1=\frac{1+\xi}{2},\, \phi_k=L_k(\xi)-L_{k-2}(\xi),\quad 
2 \leq  k \leq N.
\endeq 
The one-to-one correspondence between the Lagrangian basis and the $p$ representation yields 
\beq
v(\xi_i)=\sum^N_{k=0}{\hat v_k} \phi_k(\xi_i)\label{eq:oto}
\endeq
which in matrix notation reads
\beq
{\bf v}=\mbox{\boldmath $\Phi$}{\hat {\bf v}}.
\endeq
The filter operation is performed in spectral space through a diagonal matrix ${\bf T}$ with components such that
\beq
T_k=\frac{1}{(1+(k/k_c)^2)} \label{eq:T_k}
\endeq
where the cut-off value  $k_c$ corresponds to $T_k=1/2$. The entire filtering process for a $1D$ problem is given by
\beq
{\overline {\bf v}} =G\star {\bf v}=\mbox{\boldmath $\Phi$}\, {\bf T} \, \mbox{\boldmath $\Phi$}^{-1}\, {\bf v}
\endeq
The extension to three-dimensional is trivial by the matrix tensor product properties.

\indent
The transfer function $T_k$ of Eq. (\ref{eq:T_k}) corresponds to a standard low-pass filter required when using any spectral method \cite{boyd98}.
\subsection{The filter length}
For a $1D$ problem using the spectral element method, a common choice \cite{karam00} of filter length is
\beq
\Delta=\frac{s}{p},
\endeq
where $s$ is the element size and $p$ the highest polynomial degree in the spectral decomposition Eq. (\ref{eq:oto}) that is the closest to the frequency $k_c$
\beq
p=k,\quad  \mbox{such that}\, \inf_k (\mid k-k_c \mid) <1/2,\quad k=0, \ldots, N.
\endeq
We notice that the filter length decreases when the element is refined. The $3D$ formula for rectilinear elements is
\beq
\Delta(x,y,z)=(\Delta_1(x)\Delta_2(y)\Delta_3(z))^{1/3}=\left(\frac{s_1}{p_1}\frac{s_2}{p_2}\frac{s_3}{p_3}\right)^{1/3}.
\endeq
\section{The lid-driven cubical cavity problem}
The lid-driven cavity presents although the geometry is simple, complex physical phenomena. As in this case, we have no homogeneous direction, the presence of side walls confining the full flow modifies the flow patterns and the route to turbulence. For the description of the physics among the abundant literature, we refer the reader to Koseff and Street \cite{koseff84}, Jordan and Ragab \cite{jordan94}, Leriche and Gavrilakis \cite{leriche99}, and Shankar and Deshpande \cite{shankar00}.

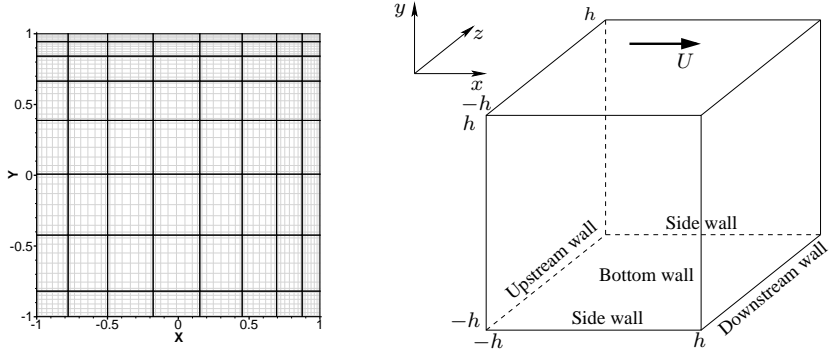
\begin{figure}[ht]
 \begin{center}
    \input{cav3d_simpl.pstex_t}
    \caption{$(a)$ (left) Spectral element grid in the mid-plane $z=0$ and $(b)$ (right) Sketch of the geometry of the lid-driven cubical cavity.}
    \label{geom_cav}
  \end{center}
\end{figure}

Figure \ref{geom_cav}.b shows the cubical cavity. The flow motion is induced by the top lid that moves in the $x$-direction with a constant unit velocity $U_0=1$. The Reynolds number is consequently $\text{Re}=U_02h/\nu$. We will essentially address the case of the flow at $\text{Re}=12'000$. The kinetic energy is provided to the flow by the shear stress at the top lid through viscous diffusion. The amplitude of the Reynolds stress below the lid is negligible indicating that the flow under the lid is mainly laminar but transient. The momentum transfer from the lid induces a region of strong pressure in the upper corner of the downstream wall as the flow, mainly horizontal prior the corner, has to change direction and moves vertically downwards. This sharp turn dissipates energy in that region. Along the downstream wall the plunging flow behaves like a jet with a variable thickness. Near the symmetry plane the jet thickness  is reduced while it increases away from this plane. This jet, laminar and unsteady at the very beginning,  separates from the cavity wall at mid-height and grows as two elliptical jets on both sides of the symmetry plane. They hit the bottom wall where they produce turbulence. This turbulence is convected away by the main central vortex towards the upstream wall where the flow slows down and relaminarizes during the fluid rise.

In order to resolve the boundary layers along the lid and the downstream wall, the spectral elements are unevenly distributed as can be seen in Figure \ref{geom_cav}.a. The spatial discretization has $E_x=E_y=E_z=8$ elements in the three space directions with $N_x=N_y=N_z=8$ polynomial degree. The spectral element calculation has two times less points per space direction than the DNS of Leriche-Gavrilakis \cite{leriche99} who employed a $129^3$ Chebyshev discretization.

As far as the velocity imposed on the lid is concerned, the unit velocity induces severe discontinuities along the top edges. In order to remove these defects we use a high degree polynomial as Leriche and Gavrilakis did 
\beq
u= \left[1-\left(x/h\right)^{18}\right]^2 \left[1-\left(z/h\right)^{18}\right]^2   ,\, v=w=0.\label{eq:bclid}
\endeq
No-slip conditions are applied to the other walls.

\indent 
Both nodal and modal filters were used in our LES computations; the former with $M=N-2$ to stabilize the velocity field at each time step and the latter with $k_c=N-2$ to filter the highest modes in the Legendre space. The computations are particularly sensitive to the values of $M$ and $k_c$; smaller values will affect spectral convergence whereas higher values will have very little effect on the smallest scales of the problem.

\indent
The dynamic constant $C_{\text{d}}$ produced has high values in the regions of high velocity gradients. Its maximum value fluctuates around 0.25 with locally some negative values which are eliminated by clipping.

\indent
The time step is chosen as $\Delta t=10^{-3}$ and the complete simulation comprises 201'000 iterations leading to a total effective simulation time of 201 time units. The reference results are the DNS data of Leriche \cite{lerichephd} and the experimental ones from Koseff and Street \cite{koseff84}, corresponding to 1'000 and 145.5 time units respectively. In the cavity flow, the average is obtained by time averaging.

\begin{figure}[ht]
  \begin{center}
    \includegraphics[width=0.99\textwidth, keepaspectratio=true]{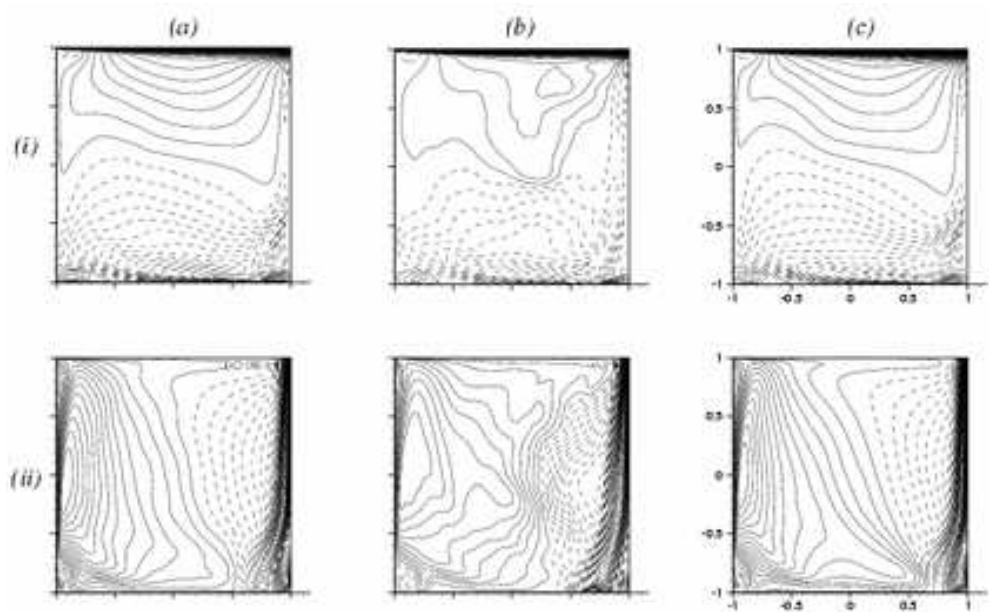}
    \caption{Contours of the $x$-component of the mean velocity field $\langle U \rangle$ (row ($i$)) and of the $y$-component $\langle V \rangle$ (row ($ii$)) in the mid-plane $z=0$ for column $(a)$ the LES, column $(b)$ the under-resolved DNS, column $(c)$ the DNS from Leriche and Gavrilakis \cite{leriche99}.}
    \label{mean_profiles}
  \end{center}
\end{figure}

The results of an under-resolved DNS performed on the coarse grid made of 65$^3$ grid points and designed for the LES are presented on Fig. \ref{mean_profiles}.b. These results are compared to the same results for the LES (Fig. \ref{mean_profiles}.a) and for the DNS (Fig. \ref{mean_profiles}.c). A qualitative comparisons of the three sets of mean-velocity contours for $\langle U \rangle$ and $\langle V\rangle$ allows to conclude to the failure of the under-resolved DNS to reproduce the physics involved in this problem. As expected the resolved-scale spectrum is smaller than the full-scale spectrum for this turbulent flow and therefore necessitates to model the under-resolved scales as it is done in our LES (Fig. \ref{mean_profiles}.a).

\begin{figure}[ht]
  \begin{center}
    \input{uvrms_xy.pstex_t}
    \caption{Profiles of the rms fluctuations $u'$ ($(a)$ and $(b)$), $v'$ ($(c)$ and $(d)$) in the mid-plane $z=0$, along the lines $x=0$ ($(b)$ and $(d)$) and $y=0$ ($(a)$ and $(c)$), for the LES, the DNS from Leriche and Gavrilakis \cite{leriche99} and the experimental data from Koseff and Street \cite{koseff84}.}
    \label{uvrms_xy}
   \end{center}
\end{figure}
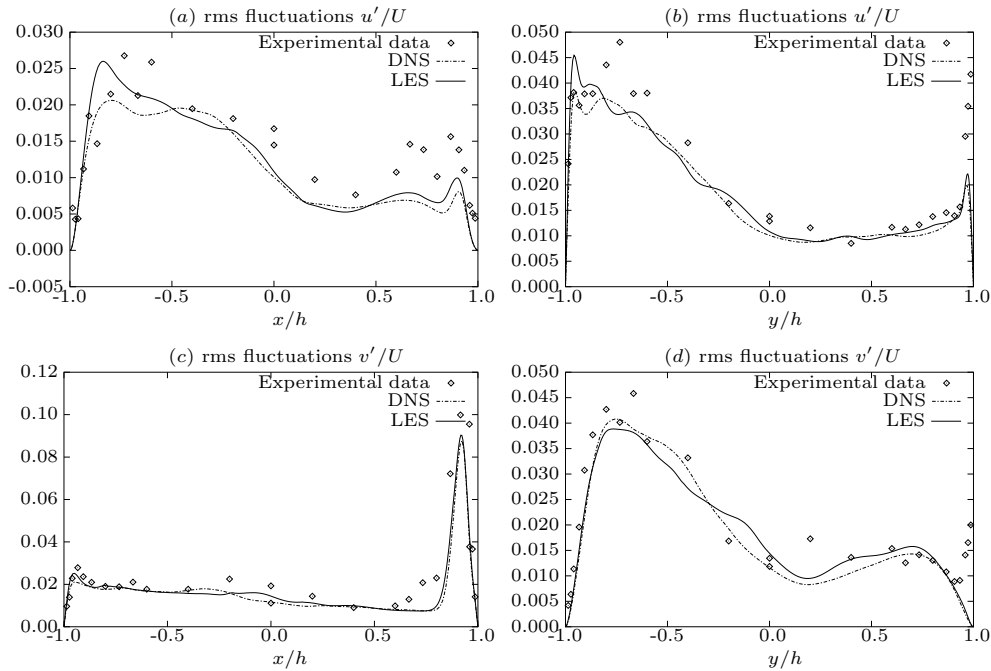

From a more quantitative point-of-view, the $x$-component $u'$ and the $y$-compo-~nent $v'$ of the rms fluctuations of the velocity field have been computed along two $1D$ lines $x=0$ and $y=0$ in the mid-plane $z=0$. Results for the LES are shown on Figure \ref{uvrms_xy} together with the DNS and experimental results, proving the excellent performance of our SGS modelling. The differences of the three sets of data in terms of peak amplitude observed can be explained by the three different ensemble averaging used as proved by Leriche \cite{lerichephd}. A larger ensemble averaging tends to reduce the amplitude of the different peaks.

\section{Conclusion and future studies}
A large-eddy simulation of the three-dimensional lid-driven cubical cavity flow using the spectral element method has been presented. The treatment of the subgrid-scales relies on a dynamic model for the eddy viscosity. This LES has been carried out on a parallel architecture with a relatively coarse grid and numerical results appeared to be extremely close to the DNS and experimental results available in the literature. Moreover the results of the under-resolved DNS on the coarse grid are far from providing any insight into the physics of the problem.

Our next goal is to perform the same LES but with other treatments of the subgrid-scales and determine the most efficient model when the simulation are based on the spectral element method.
\section{Acknowledgements}
This research is being partially funded by a Swiss National Science Fundation Grant (No. 200020-101707), whose support is gratefully acknowledged.

\end{document}

%% file: cav3d_simpl.pstex_t
\begin{picture}(0,0)%
\includegraphics[width=5cm, keepaspectratio=true]{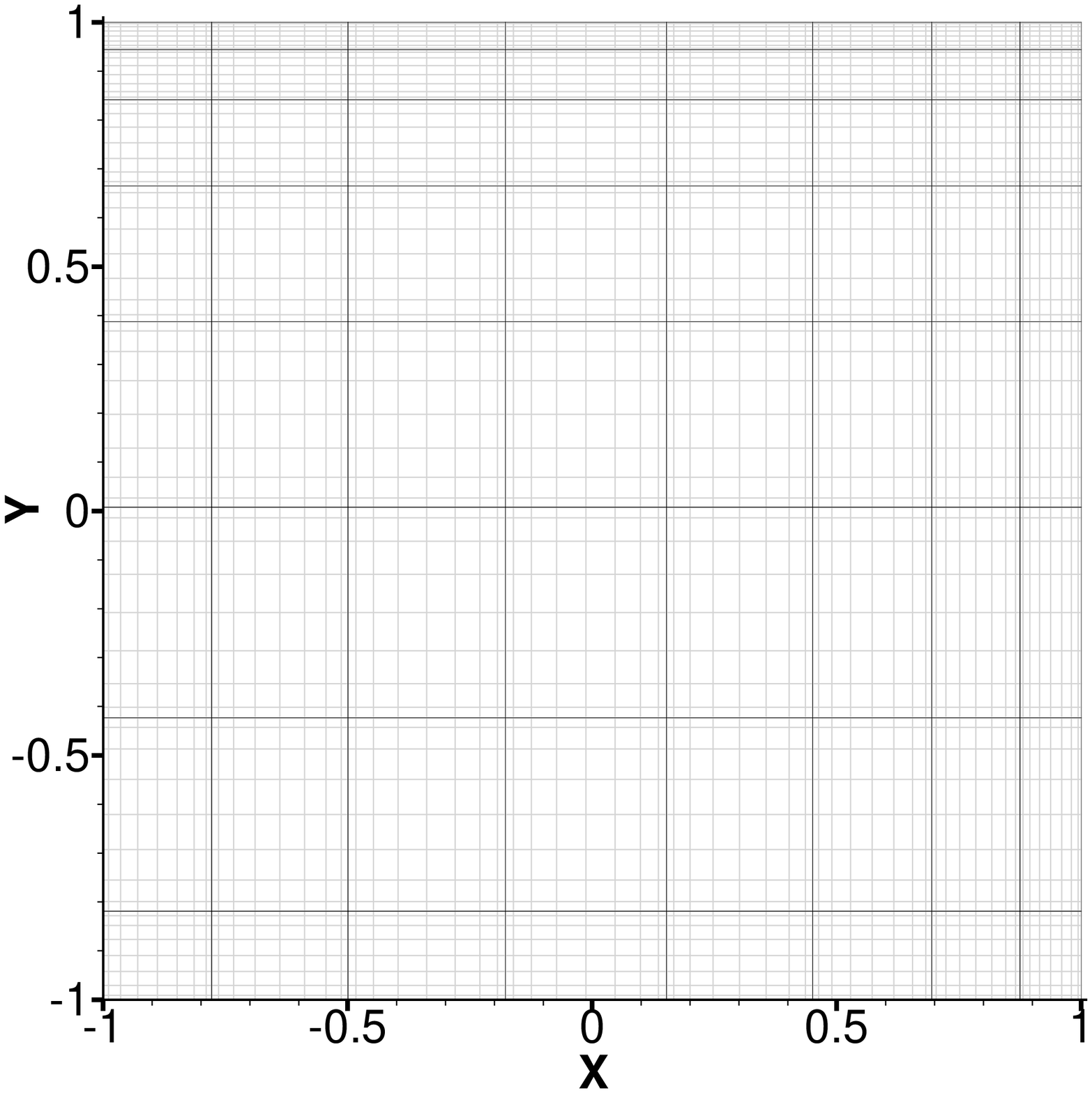}%
\end{picture}%
\begin{picture}(0,0)%
\includegraphics{cav3d_simpl.pstex}%
\end{picture}%
\setlength{\unitlength}{1973sp}%
\begingroup\makeatletter\ifx\SetFigFont\undefined%
\gdef\SetFigFont#1#2#3#4#5{%
  \reset@font\fontsize{#1}{#2pt}%
  \fontfamily{#3}\fontseries{#4}\fontshape{#5}%
  \selectfont}%
\fi\endgroup%
\begin{picture}(10707,4458)(-3328,-5143)
\put(2812,-1840){\makebox(0,0)[lb]{\smash{\SetFigFont{8}{9.6}{\rmdefault}{\mddefault}{\updefault}{\color[rgb]{0,0,0}$x$}%
}}}
\put(2840,-1199){\makebox(0,0)[lb]{\smash{\SetFigFont{8}{9.6}{\rmdefault}{\mddefault}{\updefault}{\color[rgb]{0,0,0}$z$}%
}}}
\put(1851,-877){\makebox(0,0)[lb]{\smash{\SetFigFont{8}{9.6}{\rmdefault}{\mddefault}{\updefault}{\color[rgb]{0,0,0}$y$}%
}}}
\put(5401,-1561){\makebox(0,0)[lb]{\smash{\SetFigFont{8}{9.6}{\rmdefault}{\mddefault}{\updefault}{\color[rgb]{0,0,0}$U$}%
}}}
\put(2841,-5085){\makebox(0,0)[lb]{\smash{\SetFigFont{8}{9.6}{\rmdefault}{\mddefault}{\updefault}{\color[rgb]{0,0,0}$-h$}%
}}}
\put(5603,-5077){\makebox(0,0)[lb]{\smash{\SetFigFont{8}{9.6}{\rmdefault}{\mddefault}{\updefault}{\color[rgb]{0,0,0}$h$}%
}}}
\put(2693,-2093){\makebox(0,0)[lb]{\smash{\SetFigFont{8}{9.6}{\rmdefault}{\mddefault}{\updefault}{\color[rgb]{0,0,0}$-h$}%
}}}
\put(2552,-4822){\makebox(0,0)[lb]{\smash{\SetFigFont{8}{9.6}{\rmdefault}{\mddefault}{\updefault}{\color[rgb]{0,0,0}$-h$}%
}}}
\put(2710,-2340){\makebox(0,0)[lb]{\smash{\SetFigFont{8}{9.6}{\rmdefault}{\mddefault}{\updefault}{\color[rgb]{0,0,0}$h$}%
}}}
\put(4233,-969){\makebox(0,0)[lb]{\smash{\SetFigFont{7}{8.4}{\rmdefault}{\mddefault}{\updefault}{\color[rgb]{0,0,0}$h$}%
}}}
\end{picture}

%% file: uvrms_xy.pstex_t
\begin{picture}(0,0)%
\includegraphics{uvrms_xy.pstex}%
\end{picture}%
\setlength{\unitlength}{1973sp}%
\begingroup\makeatletter\ifx\SetFigFont\undefined%
\gdef\SetFigFont#1#2#3#4#5{%
  \reset@font\fontsize{#1}{#2pt}%
  \fontfamily{#3}\fontseries{#4}\fontshape{#5}%
  \selectfont}%
\fi\endgroup%
\begin{picture}(11515,8306)(813,-8449)
\put(3507,-8386){\makebox(0,0)[lb]{\smash{\SetFigFont{7}{8.4}{\rmdefault}{\mddefault}{\updefault}{\color[rgb]{0,0,0}$x/h$}%
}}}
\put(2205,-4586){\makebox(0,0)[lb]{\smash{\SetFigFont{7}{8.4}{\rmdefault}{\mddefault}{\updefault}{\color[rgb]{0,0,0}$(c)$ rms fluctuations $v'/U$}%
}}}
\put(9732,-8386){\makebox(0,0)[lb]{\smash{\SetFigFont{7}{8.4}{\rmdefault}{\mddefault}{\updefault}{\color[rgb]{0,0,0}$y/h$}%
}}}
\put(8430,-4586){\makebox(0,0)[lb]{\smash{\SetFigFont{7}{8.4}{\rmdefault}{\mddefault}{\updefault}{\color[rgb]{0,0,0}$(d)$ rms fluctuations $v'/U$}%
}}}
\put(3507,-4111){\makebox(0,0)[lb]{\smash{\SetFigFont{7}{8.4}{\rmdefault}{\mddefault}{\updefault}{\color[rgb]{0,0,0}$x/h$}%
}}}
\put(2205,-311){\makebox(0,0)[lb]{\smash{\SetFigFont{7}{8.4}{\rmdefault}{\mddefault}{\updefault}{\color[rgb]{0,0,0}$(a)$ rms fluctuations $u'/U$}%
}}}
\put(9732,-4111){\makebox(0,0)[lb]{\smash{\SetFigFont{7}{8.4}{\rmdefault}{\mddefault}{\updefault}{\color[rgb]{0,0,0}$y/h$}%
}}}
\put(8430,-311){\makebox(0,0)[lb]{\smash{\SetFigFont{7}{8.4}{\rmdefault}{\mddefault}{\updefault}{\color[rgb]{0,0,0}$(b)$ rms fluctuations $u'/U$}%
}}}
\put(5482,-5389){\makebox(0,0)[rb]{\smash{\SetFigFont{7}{8.4}{\familydefault}{\mddefault}{\updefault}LES}}}
\put(5493,-5139){\makebox(0,0)[rb]{\smash{\SetFigFont{7}{8.4}{\familydefault}{\mddefault}{\updefault}DNS}}}
\put(813,-7973){\makebox(0,0)[rb]{\smash{\SetFigFont{7}{8.4}{\familydefault}{\mddefault}{\updefault} 0.00}}}
\put(813,-7440){\makebox(0,0)[rb]{\smash{\SetFigFont{7}{8.4}{\familydefault}{\mddefault}{\updefault} 0.02}}}
\put(813,-6906){\makebox(0,0)[rb]{\smash{\SetFigFont{7}{8.4}{\familydefault}{\mddefault}{\updefault} 0.04}}}
\put(813,-6373){\makebox(0,0)[rb]{\smash{\SetFigFont{7}{8.4}{\familydefault}{\mddefault}{\updefault} 0.06}}}
\put(813,-5840){\makebox(0,0)[rb]{\smash{\SetFigFont{7}{8.4}{\familydefault}{\mddefault}{\updefault} 0.08}}}
\put(813,-5306){\makebox(0,0)[rb]{\smash{\SetFigFont{7}{8.4}{\familydefault}{\mddefault}{\updefault} 0.10}}}
\put(813,-4773){\makebox(0,0)[rb]{\smash{\SetFigFont{7}{8.4}{\familydefault}{\mddefault}{\updefault} 0.12}}}
\put(888,-8098){\makebox(0,0)[b]{\smash{\SetFigFont{7}{8.4}{\familydefault}{\mddefault}{\updefault}-1.0}}}
\put(2188,-8098){\makebox(0,0)[b]{\smash{\SetFigFont{7}{8.4}{\familydefault}{\mddefault}{\updefault}-0.5}}}
\put(3488,-8098){\makebox(0,0)[b]{\smash{\SetFigFont{7}{8.4}{\familydefault}{\mddefault}{\updefault} 0.0}}}
\put(4788,-8098){\makebox(0,0)[b]{\smash{\SetFigFont{7}{8.4}{\familydefault}{\mddefault}{\updefault} 0.5}}}
\put(6088,-8098){\makebox(0,0)[b]{\smash{\SetFigFont{7}{8.4}{\familydefault}{\mddefault}{\updefault} 1.0}}}
\put(5488,-4910){\makebox(0,0)[rb]{\smash{\SetFigFont{7}{8.4}{\familydefault}{\mddefault}{\updefault}Experimental data}}}
\put(11707,-5389){\makebox(0,0)[rb]{\smash{\SetFigFont{7}{8.4}{\familydefault}{\mddefault}{\updefault}LES}}}
\put(11718,-5139){\makebox(0,0)[rb]{\smash{\SetFigFont{7}{8.4}{\familydefault}{\mddefault}{\updefault}DNS}}}
\put(7113,-7973){\makebox(0,0)[rb]{\smash{\SetFigFont{7}{8.4}{\familydefault}{\mddefault}{\updefault} 0.000}}}
\put(7113,-7653){\makebox(0,0)[rb]{\smash{\SetFigFont{7}{8.4}{\familydefault}{\mddefault}{\updefault} 0.005}}}
\put(7113,-7333){\makebox(0,0)[rb]{\smash{\SetFigFont{7}{8.4}{\familydefault}{\mddefault}{\updefault} 0.010}}}
\put(7113,-7013){\makebox(0,0)[rb]{\smash{\SetFigFont{7}{8.4}{\familydefault}{\mddefault}{\updefault} 0.015}}}
\put(7113,-6693){\makebox(0,0)[rb]{\smash{\SetFigFont{7}{8.4}{\familydefault}{\mddefault}{\updefault} 0.020}}}
\put(7113,-6373){\makebox(0,0)[rb]{\smash{\SetFigFont{7}{8.4}{\familydefault}{\mddefault}{\updefault} 0.025}}}
\put(7113,-6053){\makebox(0,0)[rb]{\smash{\SetFigFont{7}{8.4}{\familydefault}{\mddefault}{\updefault} 0.030}}}
\put(7113,-5733){\makebox(0,0)[rb]{\smash{\SetFigFont{7}{8.4}{\familydefault}{\mddefault}{\updefault} 0.035}}}
\put(7113,-5413){\makebox(0,0)[rb]{\smash{\SetFigFont{7}{8.4}{\familydefault}{\mddefault}{\updefault} 0.040}}}
\put(7113,-5093){\makebox(0,0)[rb]{\smash{\SetFigFont{7}{8.4}{\familydefault}{\mddefault}{\updefault} 0.045}}}
\put(7188,-8098){\makebox(0,0)[b]{\smash{\SetFigFont{7}{8.4}{\familydefault}{\mddefault}{\updefault}-1.0}}}
\put(8469,-8098){\makebox(0,0)[b]{\smash{\SetFigFont{7}{8.4}{\familydefault}{\mddefault}{\updefault}-0.5}}}
\put(9751,-8098){\makebox(0,0)[b]{\smash{\SetFigFont{7}{8.4}{\familydefault}{\mddefault}{\updefault} 0.0}}}
\put(11032,-8098){\makebox(0,0)[b]{\smash{\SetFigFont{7}{8.4}{\familydefault}{\mddefault}{\updefault} 0.5}}}
\put(12313,-8098){\makebox(0,0)[b]{\smash{\SetFigFont{7}{8.4}{\familydefault}{\mddefault}{\updefault} 1.0}}}
\put(11713,-4910){\makebox(0,0)[rb]{\smash{\SetFigFont{7}{8.4}{\familydefault}{\mddefault}{\updefault}Experimental data}}}
\put(7113,-4773){\makebox(0,0)[rb]{\smash{\SetFigFont{7}{8.4}{\familydefault}{\mddefault}{\updefault} 0.050}}}
\put(5482,-1114){\makebox(0,0)[rb]{\smash{\SetFigFont{7}{8.4}{\familydefault}{\mddefault}{\updefault}LES}}}
\put(5493,-864){\makebox(0,0)[rb]{\smash{\SetFigFont{7}{8.4}{\familydefault}{\mddefault}{\updefault}DNS}}}
\put(888,-3698){\makebox(0,0)[rb]{\smash{\SetFigFont{7}{8.4}{\familydefault}{\mddefault}{\updefault}-0.005}}}
\put(888,-3241){\makebox(0,0)[rb]{\smash{\SetFigFont{7}{8.4}{\familydefault}{\mddefault}{\updefault} 0.000}}}
\put(888,-2784){\makebox(0,0)[rb]{\smash{\SetFigFont{7}{8.4}{\familydefault}{\mddefault}{\updefault} 0.005}}}
\put(888,-2327){\makebox(0,0)[rb]{\smash{\SetFigFont{7}{8.4}{\familydefault}{\mddefault}{\updefault} 0.010}}}
\put(888,-1869){\makebox(0,0)[rb]{\smash{\SetFigFont{7}{8.4}{\familydefault}{\mddefault}{\updefault} 0.015}}}
\put(888,-1412){\makebox(0,0)[rb]{\smash{\SetFigFont{7}{8.4}{\familydefault}{\mddefault}{\updefault} 0.020}}}
\put(888,-955){\makebox(0,0)[rb]{\smash{\SetFigFont{7}{8.4}{\familydefault}{\mddefault}{\updefault} 0.025}}}
\put(888,-498){\makebox(0,0)[rb]{\smash{\SetFigFont{7}{8.4}{\familydefault}{\mddefault}{\updefault} 0.030}}}
\put(963,-3823){\makebox(0,0)[b]{\smash{\SetFigFont{7}{8.4}{\familydefault}{\mddefault}{\updefault}-1.0}}}
\put(2244,-3823){\makebox(0,0)[b]{\smash{\SetFigFont{7}{8.4}{\familydefault}{\mddefault}{\updefault}-0.5}}}
\put(3526,-3823){\makebox(0,0)[b]{\smash{\SetFigFont{7}{8.4}{\familydefault}{\mddefault}{\updefault} 0.0}}}
\put(4807,-3823){\makebox(0,0)[b]{\smash{\SetFigFont{7}{8.4}{\familydefault}{\mddefault}{\updefault} 0.5}}}
\put(6088,-3823){\makebox(0,0)[b]{\smash{\SetFigFont{7}{8.4}{\familydefault}{\mddefault}{\updefault} 1.0}}}
\put(5488,-635){\makebox(0,0)[rb]{\smash{\SetFigFont{7}{8.4}{\familydefault}{\mddefault}{\updefault}Experimental data}}}
\put(11707,-1114){\makebox(0,0)[rb]{\smash{\SetFigFont{7}{8.4}{\familydefault}{\mddefault}{\updefault}LES}}}
\put(11718,-864){\makebox(0,0)[rb]{\smash{\SetFigFont{7}{8.4}{\familydefault}{\mddefault}{\updefault}DNS}}}
\put(7113,-3698){\makebox(0,0)[rb]{\smash{\SetFigFont{7}{8.4}{\familydefault}{\mddefault}{\updefault} 0.000}}}
\put(7113,-3378){\makebox(0,0)[rb]{\smash{\SetFigFont{7}{8.4}{\familydefault}{\mddefault}{\updefault} 0.005}}}
\put(7113,-3058){\makebox(0,0)[rb]{\smash{\SetFigFont{7}{8.4}{\familydefault}{\mddefault}{\updefault} 0.010}}}
\put(7113,-2738){\makebox(0,0)[rb]{\smash{\SetFigFont{7}{8.4}{\familydefault}{\mddefault}{\updefault} 0.015}}}
\put(7113,-2418){\makebox(0,0)[rb]{\smash{\SetFigFont{7}{8.4}{\familydefault}{\mddefault}{\updefault} 0.020}}}
\put(7113,-2098){\makebox(0,0)[rb]{\smash{\SetFigFont{7}{8.4}{\familydefault}{\mddefault}{\updefault} 0.025}}}
\put(7113,-1778){\makebox(0,0)[rb]{\smash{\SetFigFont{7}{8.4}{\familydefault}{\mddefault}{\updefault} 0.030}}}
\put(7113,-1458){\makebox(0,0)[rb]{\smash{\SetFigFont{7}{8.4}{\familydefault}{\mddefault}{\updefault} 0.035}}}
\put(7113,-1138){\makebox(0,0)[rb]{\smash{\SetFigFont{7}{8.4}{\familydefault}{\mddefault}{\updefault} 0.040}}}
\put(7113,-818){\makebox(0,0)[rb]{\smash{\SetFigFont{7}{8.4}{\familydefault}{\mddefault}{\updefault} 0.045}}}
\put(7113,-498){\makebox(0,0)[rb]{\smash{\SetFigFont{7}{8.4}{\familydefault}{\mddefault}{\updefault} 0.050}}}
\put(7188,-3823){\makebox(0,0)[b]{\smash{\SetFigFont{7}{8.4}{\familydefault}{\mddefault}{\updefault}-1.0}}}
\put(8469,-3823){\makebox(0,0)[b]{\smash{\SetFigFont{7}{8.4}{\familydefault}{\mddefault}{\updefault}-0.5}}}
\put(9751,-3823){\makebox(0,0)[b]{\smash{\SetFigFont{7}{8.4}{\familydefault}{\mddefault}{\updefault} 0.0}}}
\put(11032,-3823){\makebox(0,0)[b]{\smash{\SetFigFont{7}{8.4}{\familydefault}{\mddefault}{\updefault} 0.5}}}
\put(12313,-3823){\makebox(0,0)[b]{\smash{\SetFigFont{7}{8.4}{\familydefault}{\mddefault}{\updefault} 1.0}}}
\put(11713,-635){\makebox(0,0)[rb]{\smash{\SetFigFont{7}{8.4}{\familydefault}{\mddefault}{\updefault}Experimental data}}}
\end{picture}